\documentclass[aps,pre,amsfonts,amssymb,amsmath,float, twocolumn, superscriptaddress]{revtex4-1}
\usepackage{amsthm}
\usepackage{color}
\usepackage{environ}
\usepackage{graphicx}
\usepackage{hyperref}

\usepackage[normalem]{ulem}

\newcommand{\eqlabel}[1]{\label{eq:#1}}

\NewEnviron{eqsplit}[1]{\begin{subequations}\eqlabel{#1}\begin{align} \BODY \end{align}\end{subequations}}
\newcommand{\nn}{\nonumber}
\newcommand{\ms}{\ensuremath{\mathrm{ms}}} % millisecond
\newcommand{\mm}{\ensuremath{\mathrm{mm}}} % millimeter
\newcommand{\second}{\ensuremath{\mathrm{s}}} % second

\newcommand{\avg}[1]{\overline{#1}}     % (sliding) averagve
\newcommand{\bbraket}[2]{\langle\langle#1\mid#2\rangle\rangle} % long inner product
\newcommand{\bbra}[1]{\langle\langle#1\mid}\newcommand{\kket}[1]{\mid#1\rangle\rangle} % , two halves
\newcommand{\braket}[2]{\langle #1\mid #2\rangle} % short inner product 
\newcommand{\bra}[1]{\langle#1\mid}\newcommand{\ket}[1]{\mid#1\rangle} % , two halves
\newcommand{\bzero}{\boldsymbol{0}}     % zero column
\renewcommand{\d}{\mathrm{d}}           % ordinary differential
\newcommand{\dd}{\partial}              % partial differential 
\newcommand{\diag}{\mathrm{diag}}       % diagonal mx
\renewcommand{\i}{\mathrm{i}}           % imaginary unit
\newcommand{\Lap}{\Delta}               % Laplacian
\newcommand{\Real}{\mathbb{R}}          % set of reals
\newcommand{\OO}{\mathcal{O}}           % asymptotic order
\newcommand{\Zahlen}{\mathbb{Z}}        % set of integers
\DeclareMathOperator{\Heav}{H}     % Heaviside step function
\ifx\notcolor\undefined\newcommand{\notcolor}{blue}\else\fi
\newcommand{\+}[2]{\def#1{{\color{\notcolor}#2}}}
\newcommand{\1}[2]{\def#1##1{{\color{\notcolor}#2}}}
\newcommand{\2}[2]{\def#1##1##2{{\color{\notcolor}#2}}}
\newcommand{\3}[2]{\def#1##1##2##3{{\color{\notcolor}#2}}}
\+{\A}{A}\+{\B}{B}\+{\C}{C}\+{\D}{D}    % corot frame indices
\+{\Amp}{A}                             % buckling amplitude in figure 2
\+{\a}{a} \+{\b}{b} \+{\c}{c}           % lab frame indices
\+{\adj}{^\dagger}                      % operator adjoint
\+{\alp}{\chi}                          % meandering angle
\+{\alpo}{\alp}                       % .., unperturbed
\+{\apar}{a}\+{\bpar}{b}\+{\cpar}{c}    % Barkley parameters a,b,c=epsilon
\+{\bF}{\mathbf{F}}                     % reaction rates column-function
\+{\bP}{\mathbf{P}}                     % "any 2x2 matrix" in Sec A3
\+{\bQ}{\mathbf{Q}}                     % tensor of mobilities
\+{\bR}{\mathbf{R}}                     % "rotation matrix" in Sec A3
\+{\beps}{\boldsymbol{\epsilon}}        % generator of rotations
\+{\Ctr}{C}                             % meander centre
\+{\CL}{\hat{\boldsymbol{\mathcal{L}}}} % long linearized operator
\+{\dist}{d}                            % typical length scale
\+{\E}{E}                               % electric field
\+{\Ecrit}{E_{crit}}                    % .., critical value
\+{\ex}{\vec{e}_x}                      % unit vector in x direction in fig 3
\+{\ez}{\vec{e}_z}                      % unit vector in z direction in definition of $\vt$ for 2D
\+{\f}{\mathbf{f}}\+{\g}{\mathbf{g}}    % arbitrary column fields
\+{\ff}{f}                              % any function to be averaged
\+{\findex}{f}                          % phase index in sec A3
\+{\G}{\Gamma}                     % filament tension, either
\+{\Gone}{\Gamma_1}                     % filament tension, scalar component
\+{\Gtwo}{\Gamma_2}                     % filament tension, pseudoscalar component
\+{\gone}{\gamma_1}                     % filament tension, scalar component
\+{\gtwo}{\gamma_2}                     % filament tension, pseudoscalar component
\+{\HL}{\hat{\mathbf{L}}}               % short linearizer operator
\+{\HM}{{\mathbf{Q}}}               % matrix of mobilities
\+{\HP}{{\mathbf{P}}}               % diffusion matrix
\+{\Herm}{^{\mathrm H}}                 % Hermitian transposed
\+{\h}{\mathbf{h}}                      % perturbation column
\+{\heta}{\eta}                         % .., magnitude (esp. as small parameter)
\+{\j}{j}                               % index of a component; enumerator of "fiducial points" in sec B2
\2{\kron}{\delta^{#1}_{#2}}             % Kronecker delta as tensor
\+{\kur}{k}                             % filament curvature
\+{\Lz}{L_z}                            % domain height in figure 2
\+{\la}{\lambda}                        % e-val
\+{\lcs}{\epsilon}                      % 2D Levi-Civita; NB \eps^A_{\hs B} vs \eps^{\hs A}_B ?
\2{\lcsf}{{\lcs^{#1}}_{#2}}             % .. "forward" variant
\2{\lcsb}{{\lcs_{#1}}^{#2}}             % .. "backward" variant
\+{\MM}{\mathcal{M}}                    % meandering map (isometry of meandering period)
\+{\M}{M}\+{\N}{N}                      % collective coordinate indices
\2{\NC}{A^{#1}_{#2}}                    % Meander Lemma normalization const
\+{\m}{m}                               % collective coordinate index
\+{\mOmgo}{\Omega}                    % .., unperturbed
\+{\mPsi}{\Psi}                         % unperturbed meandering phase
\+{\mpsi}{\psi}\+{\mpsio}{\psi_0}       % meandering phase, its initial value
\+{\mpsip}{\psi^*} % collective coordinates at time of perturbation
\+{\mphip}{\phi^*}
\+{\xpp}{x_p^*}
\+{\ypp}{y_p^*}
\1{\c}{c^{#1}}
\+{\n}{n}                               % an integer in the Meander Lemma, collective coordinate index
\+{\romg}{\omega}                       % meandering pattern angular velocity
\+{\romgo}{\omega}                     % .., unperturbed
\+{\somg}{\omega_s}                     % .., of the spiral 
\+{\polang}{\theta}                     % polar angle in corot frame
\2{\P}{{P^{#1}}_{#2}}                   % permittivity matrix element (??)
\2{\Q}{{Q^{#1}}_{#2}}                   % perturbation matrix element
\+{\q}{\mu}                             % averaged Q components in phase lock theory
\2{\R}{{R^{#1}}_{#2}}                   % rotation tensor
\+{\Rad}{R}                            % averaged radius of the unperturbed meander flower
\+{\r}{\vec{r}}                         % position vector, lab frame
\+{\ro}{r}                            % tip traj in corot frame
\+{\rphi}{\phi}\+{\rphio}{\phi_0}       % rotation phase, its initial value
\+{\rphil}{\phi_{\ell}}                 % .., locked value
\+{\Thick}{L}                           % medium thickness
\+{\Tip}{T}                             % tip coords
\+{\To}{T}                            % meandering period, unperturbed
\+{\t}{t}                               % time, lab frame
\2{\tQ}{{\tilde{Q}\mbox{}^{#1}}_{#2}}   % "rotated" perturbation matrix element in sec A3
\2{\tP}{{\tilde{P}\mbox{}^{#1}}_{#2}}   % permittivity matrix element averaged (??)
\+{\tWW}{\widetilde{\WW}}                   % left e-funcs, tip-displaced
\+{\t}{t}                               % time, lab frame
\+{\tuu}{\tilde{\uu}}                   % the RD column field correction
\+{\uvar}{u}                            % u-field in Barkley, FK
\+{\uu}{\mathbf{u}}                     % the RD column field
\+{\uuo}{\uu_0}                         % .., unperturbed spiral
\+{\V}{V}                               % (generalised) drift velocities
\+{\VV}{\mathbf{V}}                     % right e-func
\+{\Vpar}{V_\parallel}\+{\Vperp}{V_\perp}    % drift velocity components
\+{\vvar}{v}                            % v-field in Barkley, FK
\+{\vB}{\vec{B}}                        % filament binormal vector
\+{\vE}{\vec{E}}                        % electric field vector
\+{\vN}{\vec{N}}                        % filament main normal vector
\+{\vT}{\vec{\mathcal{T}}}              % filament unit tangent vector
\+{\vo}{v} \+{\vvo}{\vec{v}}        % tip vel in corot frame
 \1{\vo}{v^{#1}}        % tip vel in corot frame
\+{\vpar}{v_\parallel}\+{\vperp}{v_\perp}    % drift velocity components
\+{\v}{\tilde{v}}                               % \(generalised) drift velocities
\+{\vnabla}{\nabla}                     %  so vector sign above it is spurioue
\+{\vV}{\vec{V}}                        % Average drift velocity
\+{\wvar}{w}                            % w-field in FK
\+{\WW}{\mathbf{W}}\+{\W}{W}            % left e-funcs, its component
\+{\X}{X}                               % centre position in index notation
\+{\Xc}{X^1} \+{\Yc}{X^2}                   % centre coords
\+{\x}{x}                               % generic position vector in index notation; Cartesian x index
\+{\xl}{x} \+{\yl}{y}                   % lab frame coords
\1{\xp}{\rho^{#1}}            % corot frame coords 
\+{\tp}{\tau}                      % corot frame time
\+{\y}{y}                               % Cartesian y index
\+{\Block}{B}                           % block function 
\+{\dt}{dt}                             % time step
\+{\dx}{dx}                             % space step
\+{\Eo}{E_0}                            % perturbation magnitude
\+{\Nx}{N_x}\+{\Ny}{N_y}\+{\Nz}{N_z}    % grid dimensions
\+{\tptb}{t_p}                          % perturbation time moment
\+{\ttwo}{t_2}                          % test time moment
\+{\tend}{t_{\rm{end}}}                          % test time moment
\+{\Deltat}{\Delta_t}                   % perturbation duration
\+{\somex}{x}                           % dummy arg in pulse function definition
\+{\someDelta}{\Delta}                  % dummy arg in pulse function definition
\1{\Xo}{X_0^{#1}}                            % init position vector coords
\+{\Xco}{X_0} \+{\Yco}{Y_0}             % .., by name

\renewcommand{\@}{\partial}             % partial differential
\providecommand{\abs}[1]{{\left\lvert #1 \right\rvert}}
\newcommand{\cc}{\textrm{c.c.}}         % complex conjugate
\newcommand{\e}{\mathrm{e}}             % nat log base
\renewcommand{\O}[1]{\mathcal{O}\!\left(#1\right)} % asymptotic order
\newcommand{\sumz}[1]{\sum\limits_{#1\in\Zahlen}}
\2{\FF}{F_{#1,#2}}                    % .., its double-Fourier coeffs
\+{\k}{k}                               % Fourier index
\+{\ki}{k'}                             % Fourier index
\+{\l}{\ell}                            % Fourier index
\+{\mpsif}{\psi_*}                      % meander phase at the reference moment
\+{\omin}{\omega_{\min}}                % minimal combination frequency
\+{\ocut}{\omega_c}                     % filter cut-off frequency
\3{\QF}{{\mbox{}^{#1}\!Q^{#2}}_{#3}}    % perturbation matrix element Fourier coefficient
\+{\rphif}{\phi_*}                      % rotation phase at the reference moment
\2{\S}{{S^{#1}}_{#2}}                   % phase correction 
\+{\tf}{t_*}                            % the reference moment
\+{\tr}{\tilde{t}}                           % time relative to the reference moment
\+{\ts}{\tau'}                          % dummy integration variable
\+{\weight}{w}                          % averaging weight function%
\2{\vF}{\mbox{}^{#1}\!v^{#2}}           % comov frame tip velocity Fourier coefficient

\newcommand{\hddel}[1]{\relax}

\begin{document}
%\title{Two regimes of filament tension for meandering scroll waves}
\title{Filament Tension and Phase-Locking of Meandering Scroll Waves}

\author{Hans Dierckx}
\affiliation{Department of Physics and Astronomy, Ghent University, 9000 Ghent, Belgium}
\author{I.V. Biktasheva}
\affiliation{Department of Computer Science, University of Liverpool, Liverpool L69 3BX, UK}
\affiliation{College of Engineering, Mathematics and Physical Sciences, University of Exeter, Exeter EX4 4QF, UK}
\author{H. Verschelde}
\affiliation{Department of Physics and Astronomy, Ghent University, 9000 Ghent, Belgium}
\author{A.V. Panfilov}
\affiliation{Department of Physics and Astronomy, Ghent University, 9000 Ghent, Belgium}
\author{V.N. Biktashev}
\affiliation{College of Engineering, Mathematics and Physical Sciences, University of Exeter, Exeter EX4 4QF, UK}

\date{\today}

\begin{abstract}
Meandering spiral waves are often observed in excitable media such as the Belousov-Zhabotinsky reaction and cardiac tissue. We derive a theory for drift dynamics of meandering rotors in general reaction-diffusion systems and apply it to two types of external disturbances: an external field and curvature-induced drift in three dimensions. We find two distinct regimes: with small filament curvature, meandering scroll waves exhibit filament tension, whose sign determines the stability and drift direction. In the regimes of strong external fields or meandering motion close to resonance, however, phase-locking of the meander pattern is predicted and observed. 
\end{abstract}

\maketitle

\begin{figure}[h] \centering
%\raisebox{2.7cm}{a)}
%[height=4.cm]
 \includegraphics[width= 0.24\textwidth]
{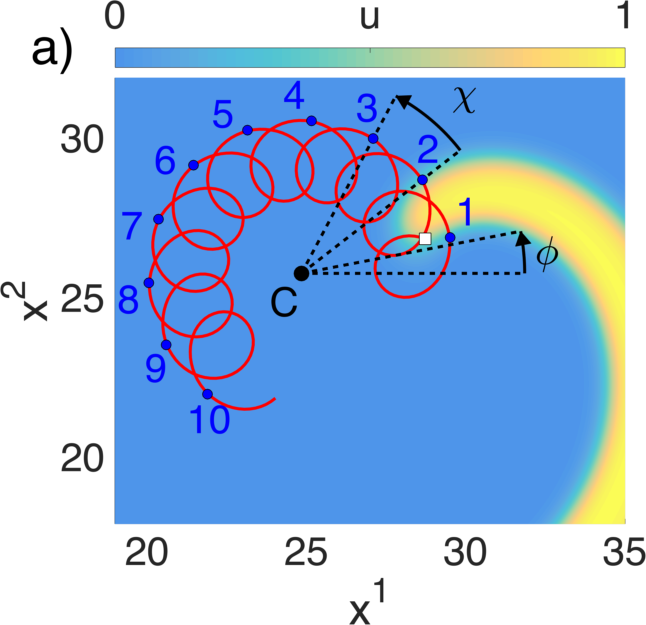} 
%\raisebox{2.7cm}{b)} 
\includegraphics[width= 0.21\textwidth]{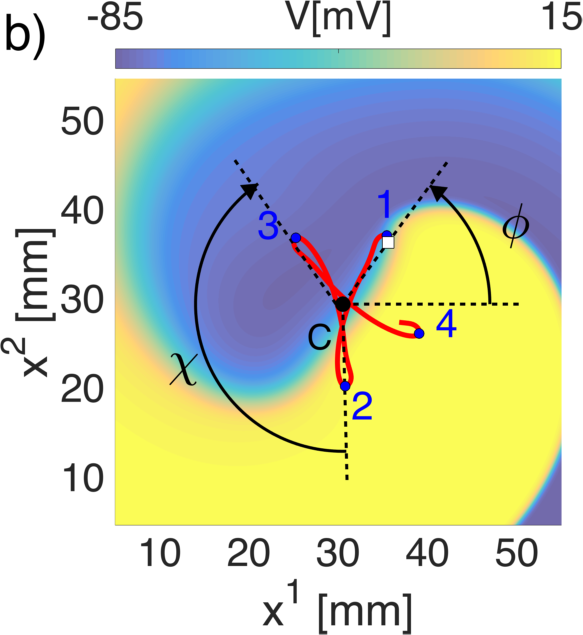}
\caption{(color online) Meandering spiral wave, with current tip position (white) and upcomping tip trajectory (red) for Barkley (a) and Fenton-Karma (b) kinetics. Successive `petals' (blue) are reached after time $\To$ and span the angle $\alpo$. 
 \label{fig:spiral}}
%\end{figure}
%\begin{figure}[bh] \centering
\raisebox{2.5cm}{a)}
\includegraphics[width=0.4\textwidth]{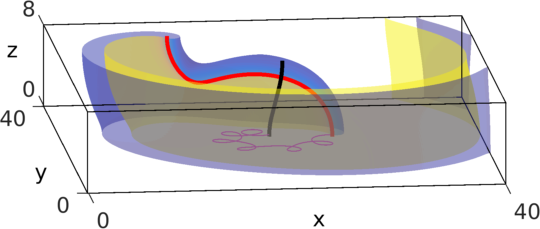} \\
\raisebox{2.5cm}{b)} 
\includegraphics[width=0.4\textwidth]{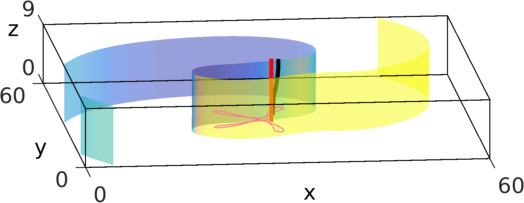}
\caption[]{(color online) Meandering 3D scroll wave evolved from a nearly straight scroll (initial filament shown in black), showing tip trajectory on the bottom surface, current wave front (blue), wave back (yellow) and filament (red). (a) For  Barkley kinetics, at $\t=30$ the filament buckles and subsequently breaks up. (b) In the FK model ($\t=950\,\ms$), the filament straightens. 
\label{fig:scroll}}
\end{figure}

\textit{Introduction.}
Rotating spiral waves are remarkable patterns which spontaneously
occur in many spatially extended systems \cite{Jahnke:1988,
  Siegert:1992, Lechleiter:1991, Gorelova:1983, Allessie:1973,
  Gray:1998, Witkowsky:1998,  Haissaguerre:2014}. In many cases, a
quasi-periodic motion of the wave pattern instead of rigid-body
rotation can be recognised from the star- or flower-like tip
trajectory, shown by red traces in Fig. \ref{fig:spiral}. These are
called `meandering spiral waves' or `modulated rotating waves'
\cite{Winfree:1991,Barkley:1994} observed in 
Belousov-Zhabotinsky (BZ) chemical reaction \cite{Winfree:1973,
  Steinbock:1993}, and in cardiac tissue experiment
\cite{Haissaguerre:2014, Yamazaki:2012} and numerical simulations
\cite{Efimov:1995, Courtemanche:1998, 
%Courtemanche:1998, Priebe:1998, 
Biktashev:1996model, Fenton:1998, BuenoOrovio:2008}. 

The understanding of the excitation patterns exhibited by circular-core spirals in 2D, and scroll waves in 3D, has much benefited
from the analysis of their motion in terms of `phase singularities',
i.e. instantaneous rotation centers for the spirals, Fig. \ref{fig:spiral},
  and `filaments', Fig. \ref{fig:scroll}, for  the scrolls
  \cite{Keener:1988, Biktashev:1994, Clayton:2005}. 
 Much of the theory of meandering spiral waves has been focusing on the origin of the meander bifurcation  \cite{Barkley:1992, Barkley:1994,
  Biktashev:1996, Golubitsky:1997},
which produces epi- or hypocyclodial motion of a spiral tip, as shown in Fig. \ref{fig:spiral}a.
However, meandering spirals with `linear' cores, as in Fig. \ref{fig:spiral}b, may be the building blocks of ventricular
fibrillation, which motivated recent work to calculate their leading eigenmodes \cite{Marcotte:2015, Marcotte:2016,
  Dierckx:2017}. In this Letter, we derive equations of motion
  for biperiodic meandering 2D spirals and 3D scroll waves, without
  restriction to a particular shape of meander.

In 3D, it
has been shown that the filament of a \emph{circular-core} scroll wave
is characterized by its `tension' ($\gone$), which
depends on the medium parameters: $\gone <0$ leads
to ever-growing filaments \cite{Panfilov:1987, Biktashev:1994} if the medium is thick enough \cite{Dierckx:2012}, resulting in a
turbulent, fibrillation-like state, while $\gone>0$ leads to shrinking of scroll rings, so that only
filaments connecting opposite medium boundaries persist. Fig. \ref{fig:scroll} shows
similar behaviour for \emph{meandering} scroll waves. 
However, the applicability of the concept of filament tension to
meandering scrolls has so far been a conjecture rather than fact. In this Letter, we will show when this is indeed true, and when it is not.

\textit{Methods.} We investigate spiral-shaped solutions to the
reaction-diffusion (RD) system in 2 and 3 spatial dimensions under a
small spatiotemporal perturbation $\h$:
\begin{eqnarray}
  \dd_\t \uu (\r,\t ) = \HP \Lap \uu(\r,\t ) +
  \bF(\uu(\r,\t )) + \h(\r,\t ) , \label{RDE}
\end{eqnarray}
where $\uu$ is a column matrix of state variables. Eq. \eqref{RDE} describes both BZ-like chemical systems and models of cardiac tissue, depending on the choice of the diffusion constants $\HP$ and reaction kinetics $\bF(\uu)$.
We consider two different kinetics models: Barkley model
  \cite{Barkley:1991}  $\uu = [\uvar,\vvar]^T$, 
$\bF =
[\cpar^{-1}\uvar(1-\uvar)\left(\uvar-\frac{\vvar+\bpar}{\apar}\right),\uvar-\vvar]^T$,
$\HP= \diag(1,0)$, $\apar=0.58, \bpar=0.05, \cpar=0.02$,
Fig.~\ref{fig:spiral}a; and the Fenton-Karma (FK) cardiac
tissue model with guinea pig (GP) parameters \cite{Fenton:1998}, where
$\uu=[\uvar,\vvar,\wvar]^T$, %
$\HP=\diag(0.1,0,0)\,\mm^2/\ms$, Fig.~\ref{fig:spiral}b. 

 Before presenting our formalism, we recall the classical view on meander in terms of the transition of
 an unperturbed ($\h=\bzero$) spiral from rigid to biperiodic rotation
  via an equivariant Hopf bifurcation\cite{Barkley:1992,
    Barkley:1994, Biktashev:1996, Golubitsky:1997, Nikolaev:1999}.
  Let $\somg$ be the angular velocity of a rigidly rotating spiral.
  In a frame of reference rotating with $\somg$,
  this spiral is a stationary solution. The corresponding linearised
  problem always has eigenvalues on the imaginary axis, due 
  to the rotational ($\la_{R}=0$) and translational
  ($\la_{T}=\pm \i \somg$) Euclidean symmetry of the plane.  If under parameter change 
  another pair of eigenvalues crosses the imaginary axis at
  $\la_{H}=\pm \i \mOmgo$, the solution is becomes time-periodic
  with period $\To = 2\pi/\mOmgo$ in the rotating frame. The progression of the
  solution along this cycle can be labeled by the meander phase
  $\mpsi$; by definition, in the absence of
  perturbation, $\dd_\t \mpsi = \mOmgo$.  %In the lab frame, this solution
 % is a biperiodic meandering spiral. %

To describe meandering spirals without relying on the proximity of the Hopf bifurcation, we note that in the lab frame, a meandering spiral is a relative periodic orbit, meaning that after the time
  $\To$ the solution returns to the same state up to an orientation-preserving isometry $\MM$ of the
  plane~\cite{Barkley:1994,Biktashev:1996,LeBlanc:2000,Foulkes:2010}. Except for the resonant case, which falls outside our present scope, $\MM$ is rotation by an angle $\alpo$ around
    a point C. After iterating $\MM$, 
      the point C emerges as the
      centre of the meander pattern, see
    Fig.~\ref{fig:spiral}. By construction, $\alpo$ is defined up to
  an integer number of full rotations. 
  In terms of the classical approach, we can write
    $\alpo=\somg\To+2\pi\n$, for $\n\in\mathbb{N}$.
    For our formalism, the exact choice of $\alpo$ is not of
    principal importance.  We find it convenient to demand
    $\abs{\alpo}<\pi$, and define $\alpo$ as the (smallest) angle between
    consecutive `petals' of the tip path, see Fig.~\ref{fig:spiral}.
  Correspondingly, we consider a frame of reference rotating
  around C with
  $\romg=\alpo/\To$, in which the solution is also $\To$-periodic.
  As before, we define $\mOmgo=2\pi/\To$ and
    $\dd_\t\mpsi=\mOmgo$. 
   Note that this formalism equally holds for both cases shown in Fig.~\ref{fig:spiral}. 

Let angle $\rphi(\t)$ characterize the orientation of the steadily rotating frame (Fig.~\ref{fig:spiral});  by definition $\dd_\t \rphi = \romgo$ for the unperturbed spiral. 
    The
   transformation between the lab frame coordinates $\x^\a$ and
   the rotating frame coordinates $\xp\A$ is then 
   $\xp\A=\R\A\a(\rphi)\left(\x^\a-\X^\a\right)$, where
   %$\R\A\a(\rphi)= \kron\A\a\cos\rphi  +
%  \lcsf\A\a \sin\rphi$
 is the rotation matrix over an angle $\rphi$. 
  In the rotating frame, Eq.~\eqref{RDE} becomes
 % \begin{linenomath*}
    \begin{equation}
      \dd_{\tp}\uu = \HP \Lap \uu + \romgo \dd_\polang \uu %+ \vo\A \dd_\A \uu 
+ \bF(\uu).   \label{RDE-comov}  
    \end{equation}
 % \end{linenomath*}
  %
Here $\tp$ is time in that frame, 
$\Lap$ is the Laplacian in the $(\xp1,\xp2)$ plane and
 $\polang$ is the polar angle in it,
 i.e. $\dd_\polang=\lcsb\A\B \xp\A\dd_\B$. 
The unperturbed meandering spiral wave solution $\uu_0(\xp1,\xp2,\mpsi)$ to Eq. \eqref{RDE-comov} 
is $2\pi$-periodic in $\mpsi$ and satisfies
%  \begin{linenomath*}
    \begin{equation}
      \HP \Lap \uuo + \romgo \dd_\polang \uuo - \mOmgo \dd_\mpsi
      \uuo + \bF(\uuo) = \bzero . \label{equ0}
    \end{equation}
  % \end{linenomath*}

In what follows, we perform a standard perturbation technique  used before to derive drift laws for
circular-core spiral and scroll waves  \cite{Keener:1988, Biktashev:1994,  Biktashev:1995b,  Verschelde:2007}. 
This involves linearization of Eq.~\eqref{equ0} 
on $\uuo$, after which the drift caused by
  a perturbation $\h$ will be given by its projection onto the symmetry eigenmodes.

The linear operator $\CL$ associated with
Eq. \eqref{equ0} is
%  to~\eqref{RDEcomov}: 
%\begin{linenomath*}
  \begin{equation}
    \CL =  \HL - \mOmgo \dd_\mpsi,  
    \qquad
    \HL =  \HP \Lap + \romgo \dd_\polang %+ \vo\A \dd_\A
 + \bF'(\uuo).  \label{defLL}
  \end{equation}
% \end{linenomath*}
%
The operator $\HL$ is the same as used for the
circular-core case \cite{Keener:1988, Biktashev:1994,
  Verschelde:2007, Biktasheva:2009}.  By differentiating Eq. \eqref{equ0} with
respect to $\xp1,\xp2,\polang$ and $\mpsi$, we find the four critical
eigenmodes: 
$ \VV_\pm = - \frac{1}{2} \left(\dd_{1} \uuo \pm \i \dd_{2} \uuo
\right),$ $\VV_\mpsi = -\dd_\mpsi \uuo$,
$\VV_\rphi = - \dd_\polang \uuo %+ \vo\A \dd_\A \uuo / \romgo)
$, where
%\begin{align}
$  \CL \VV_\pm = \pm \i \romgo \VV_\pm, 
\CL \VV_\mpsi = \bzero,  % \label{cart}
\CL \VV_\rphi = \bzero. $
%\end{align}
Modes $\VV_\pm$ correspond to translations, $\VV_\rphi$ to
rotations and $\VV_\mpsi$ to shifts in time.

To perform projections onto the symmetry modes, we use the inner
  products
\begin{equation}
\braket{\f}{\g} = \iint_{\Real^2}  \f\Herm \g\, \d\xp1\d\xp2,\  
%\label{inner1} 
\bbraket{\f}{\g} = \int\limits_0^{2\pi} \braket{\f}{\g}\, \frac{\d\mpsi}{2\pi}. %, \label{inner2} 
\label{inner} 
\end{equation} 
The projectors $\WW^\M$, also called `response functions (RFs)' in
this context \cite{Biktasheva:2003}, are the critical eigenfunctions of the adjoint
operator $\CL\adj$, 
\begin{align}
  \CL\adj =  %\HP\Herm \Lap - \romgo \dd_\polang + \bF'\Herm(\uuo) + \mOmgo \dd_\mpsi = 
 \HL\adj + \mOmgo \dd_\mpsi,
            \quad
           \HL\adj =  \HP\Herm \Lap - \romgo \dd_\polang %- \vo\A \dd_\A 
+ \bF'\Herm(\uuo),  %\nn \\
\end{align}
such that $\CL\WW^\pm = \mp \i \romgo \WW^\pm$, %
$\CL\WW^\mpsi = \bzero$, % \label{cart} 
$\CL\WW^\rphi = \bzero$. They are $2\pi$-periodic in $\mpsi$, and
can be normalized as
\begin{equation}
  \bbraket{\WW^\M}{\VV_\N} = \kron\M\N \qquad (\M,\N\in \{+,-,\rphi,\mpsi\}). \label{orthobb}
\end{equation}
The biorthogonality property~\eqref{orthobb} is not practical 
since it involves averaging over $\mpsi$. 
However, although generally an inner product $\braket{\f}{\g}$
  depends on $\mpsi$, for products of eigenfunctions of $\CL$ 
  the following \textit{Meander Lemma} holds (see also \cite{Foulkes:2010, Marcotte:2016}): 
\begin{equation}
  \braket{\WW^\M}{\VV_\N} = \kron\M\N 
  \quad 
  \forall\mpsi, 
  \quad 
  (\M,\N \in \{+,-,\rphi, \mpsi \})    .
  \label{lemma}
\end{equation}
% for all $\mpsi$. 
Indeed, 
\begin{align}
 \mOmgo \dd_\mpsi  &\braket{\WW^\M}{\VV_\N}  
\nn \\
=& \braket{(\HL\adj- \CL\adj)\WW^\M}{\VV_\N} +  \braket{\WW^\M}{(\CL - \HL)\VV_\N} \nn\\
=& (\la_\N - \la_\M)  \braket{\WW^\M}{\VV_\N}. 
\end{align}
If $\la_\M = \la_\N$, $\braket{\WW^\M}{\VV_\N}$ is constant
and equal to $\bbraket{\WW^\M}{\VV_\N} = \kron\M\N$. If $\la_\M \neq
\la_\N$, $\braket{\WW^\M}{\VV_\N}(\mpsi) = \NC\M\N
\exp[(\la_\N-\la_\M)\mpsi/\mOmgo]$. Since this is $2\pi$-periodic, $\i
(\la_\N - \la_\M) / \mOmgo$ needs to be integer. However, in the
non-resonant case we have $0 < |\alpo| < \pi$, whence $0
< |\romgo / \mOmgo| < 1/2$. For the critical modes, $(\la_\N-\la_\M)/
\mOmgo$ thus cannot be integer, and therefore $\NC\M\N =0$ and
$\braket{\WW^\M}{\VV_\N}  = 0$.  \qed 

As a corollary, the instant orthogonality \eqref{lemma} also holds for the Cartesian
basis of eigenfunctions, i.e. $\M,\N \in \{ \xp1, \xp2, \rphi, \mpsi \}$ and pairs of a critical and non-critical mode. 

\textit{Results.} We are now ready to calculate how a small perturbation $\h$, say of order $\heta$, induces spiral wave drift. Still in 2D, we decompose the exact solution to \eqref{RDE-comov} as 
\begin{align}
\uu(\xp\A,\tp) = \uuo(\xp\A,\mpsi(\tp)) + \tuu(\xp\A,\tp) \label{decomp}
\end{align}
where $\tuu = \OO(\heta)$ is made unique at all times by the condition $\braket{\WW^\M}{\tuu} =0$ \cite{Verschelde:2007}. Then, we let the frame move with yet unknown perturbed velocities:
\begin{align}
\dd_\t\X^\A &= \v^\A, & \dd_\t\mpsi &= \mOmgo + \v^\mpsi, &\dd_\t\rphi =\romgo + \v^\polang \label{pert}
\end{align}
where $\v^\M = \OO(\heta)$. Inserting these into Eq. \eqref{RDE-comov} yields\begin{align}
  (\dd_{\tp} - \HL)\tuu - \sum_{\M=\xp1,\xp2,\polang,\mpsi} \v^\M \dd_\M \uuo =  \h + \OO(\heta^2). \label{master}
\end{align}
Finally, projection onto the RFs delivers: 
\begin{align}
 \dot{\X}^\M &= \vo\M + \braket{\WW^\M}{\h} + \OO(\heta^2) \label{eqh}
\end{align}
for $\M \in \{1,2,\rphi,\mpsi\}$ where $\vo\mpsi = \mOmgo$, $\vo\rphi = \romgo$ and $\vo\A=0$. 

The equation of motion \eqref{eqh} 
describes the spatial drift of the position $\Xc$, $\Yc$ of the centre of
the  meander pattern, its orientation $\rphi$ in the plane and the meander phase $\mpsi$ of the spiral. It
is a fundamental result in this work, as it captures the generic drift response of a meandering spiral wave to
small external disturbances $\h(\r,\t )$. 
Its form was stated before based on symmetry for particular cases of $\h$
\cite{Wulff:1996,Golubitsky:1997,LeBlanc:2000,LeBlanc:2002}, but without the overlap integral which is necessary to quantitatively predict spiral wave drift. 

In this work we choose to further study 
\begin{align} 
\h = \HM \vE \cdot \vnabla \uu, \label{EM}
\end{align}
which has several applications. E.g. in chemical
systems, $\uu$ is a vector of concentrations of reagents, and
Eqs. \eqref{RDE}, \eqref{EM} may describe the `electrophoretic' drift of
spiral waves in a constant electrical field $\vE$, if $\HM$ is the
diagonal matrix of electrical mobilities of the reagents.  More generically, in any RD system describing 3D scroll waves,
% with filament curvature $\kur$ and normal $\vN$, 
one can show that the effect of diffusion in 3 dimensions boils down to a perturbation of the form \eqref{EM}, with
$\HM=\HP$ and $\vE = \kur\vN$, where $\kur$ is the geometrical curvature of the scroll wave filament (i.e. the 3D extension of the rotation centres C) and $\vN$ the local normal vector to it \cite{Keener:1988}.

The resulting spiral and scroll wave dynamics can for both applications mentioned above be found by
substituting \eqref{EM} into the general law of motion
\eqref{eqh}. Here, we will assume that the RFs $\WW^\M$ are 
essentially localized (as shown numerically in \cite{Marcotte:2015, Marcotte:2016,Dierckx:2017})
within an area of size $\dist$
and that the spatial scale over
which the fields $\vE$ vary is larger than $\dist$. In the lab frame of
reference, this delivers: 
\begin{align}
\dd_\t\rphi &= \romgo + \Q\rphi\A(\mpsi) \R\A\a(\rphi) \E^\a, \nn \\
\dd_\t\mpsi &= \mOmgo + \Q\mpsi\A(\mpsi) \R\A\a(\rphi) \E^\a, \label{efd_comp2} \\
\dd_\t\X^\b &= \R\b\B(\rphi) \Q\B\A(\mpsi)\R\A\a(\rphi) \E^\a \nn
\end{align}

with $\Q\M\A = \bra{\WW^\M}\HM\ket{\dd_A \uuo}$. 
In the case where $\vE = \kur\vN$, $\HM = \HP$, system~\eqref{efd_comp2}
describes the evolution of the scroll wave filament position $\X^\b$
in every plane locally orthogonal to the filament. 
We have thus generalized Keener's law of motion
\cite{Keener:1988}. The main difference  is
that the coefficients $\Q\M\A$ depend on the meander phase~$\mpsi$. 

The law of motion \eqref{efd_comp2} can be simplified considerably 
by averaging it over several meander periods. This is most
easily seen using Fourier series:
\begin{align}
  \Q\M\A(\mpsi)&=\sumz{\k} \QF\k\M\A\,\e^{\i\k\mpsi} , \\
  \R\A\a(\rphi)&= \frac12 \left(\kron\A\a + \i\lcsf\A\a \right)\,\e^{\i\rphi} + \cc \nn
\end{align}
where {\cc} is the complex conjugate. 

The dynamics of the center of the meander flower in each cross-section perpendicular to the filament is then
\begin{align}
  \dd_\t\X^\b = \sum\limits_{\l=-2}^{2} \sumz{\k} 
  \FF\l\k^\b \e^{\i\l\rphi} \e^{\i\k\mpsi}
  + \O{\E^2} . \label{fourier}
\end{align}
We note that unless in resonance, the set 
$
  \{ \abs{\l\romgo+\k\mOmgo} \;\big|\; \l\in\{0,\pm1,\pm2\},\;\k\in\Zahlen  \}
$
has a strictly positive minimal element, say $\omin$.  Then, all non-constant terms in  \eqref{fourier} will oscillate at a
frequency of at least $\omin$. 

However, the sole constant term $\FF00^\b$ in the right-hand-side of \eqref{fourier} will induce a constant drift velocity: 
\begin{align}
   \FF00^\b
  =& ^0\Q\B\A \frac{\E^a}{4} \left(\kron\b\B + \i\lcsf\b\B\right) \left(\kron\A\a + \i\lcsf\A\a\right) +\cc  \nn \\
  =& \frac12 \avg{\Q\A\A} \E^\b +  \frac12 \lcsf\A\B \avg{\Q\B\A}  \lcsf\b\a \E^\a.\label{F00}
\end{align}
In vector notation, this result can be written as
\begin{align}
\vV = \Gone\vE + \Gtwo\vT \times \vE  
\label{eom2}
\end{align}
where $\vV$ is the net drift motion of the filament, $\vT$ is the unit tangent to the filament for 3D scroll waves, and $\vT = \ez$ for 2D spiral waves in the XY-plane. From Eq. \eqref{F00}, the drift components parallel and perpendicular to the applied external field $\vE$ are 
given by 
\begin{align}
  \Gone &=  \avg{\Q\A\A}  = \frac{1}{2} \bbra{\WW^\A} \HM \kket{\dd_\A \uuo},\label{tension} \\ 
  \Gtwo &= \frac12 \lcsf\A\B \avg{\Q\B\A} = \frac{\lcsf\A\B}{2} \bbra{\WW^\B} \HM \kket{\dd_\A \uuo}. \nn 
\end{align}

The time-averaged
equation of motion for meandering spiral waves \eqref{eom2} exhibits
the same dynamics as in the circular-core case. If $\h$ describes diffusive coupling in the third spatial dimension ($\vE = \kur\vN$, $\HM = \HP$), Eq. \eqref{eom2} happens to reduce to the
circular-core result from \cite{Biktashev:1994}:
\begin{align}
\vV = \Gone\kur\vN + \Gtwo\kur\vB,
\label{eombik}
\end{align}
where $\vN$ and $\vB$ are the normal and binormal vectors to the filament. Then, we
can interpret $\Gone$ and $\Gtwo$ as the scalar and pseudoscalar
filament tension.   
Since Eqs. \eqref{eom2} are the laws of motion for the
filament of a meandering scroll wave, it follows from
\cite{Biktashev:1994} that the period-averaged filament length
increases monotonically in time if $\Gone < 0$ and decreases if
$\Gone >0$.

\begin{figure}[tb]
\raisebox{2.5cm}{a)} \includegraphics[width=0.2\textwidth]{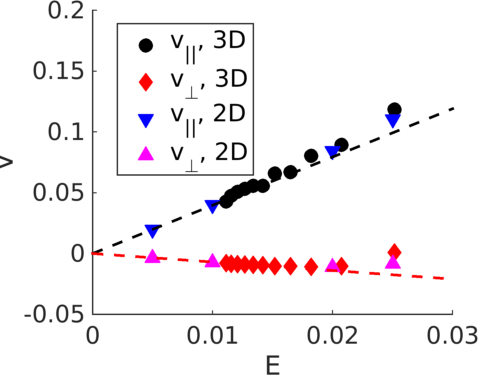} 
\raisebox{2.5cm}{b)} \includegraphics[width=0.18\textwidth]{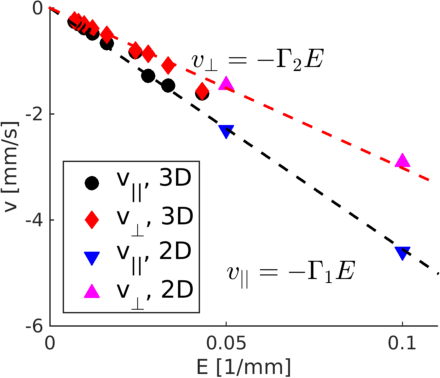} \\
%\raisebox{2.5cm}{a)} \includegraphics[width=0.23\textwidth]{fig3a.png} %{EFD_ring_Ba.png} 
%\raisebox{2.5cm}{b)} \includegraphics[width=0.2\textwidth]{fig3b.png}\\ %{EFD_ring_FK.png} \\
%\raisebox{1.5cm}{c)} \includegraphics[width=0.2\textwidth]{buckling_bif.png} 
%\raisebox{1.5cm}{d)} \includegraphics[width=0.2\textwidth]{length_FK.png} 
\caption[]{\label{fig:tension} (Color online) Comparison of  drift velocity compo\-nents with theory, in the Barkley (a) and FK model (b). `3D' refers to a scroll ring simulation with $\h = \HP \frac{1}{r}\dd_r \uu$.
}
\end{figure}

To validate our results, we have determined the coefficients
$\P\M\A(\mpsi)$
for the Barkley and FK kinetics by applying $\vE$ for a short time
interval at different values of the meander phase $\mpsi$, see
Supp. {B} for details of numerics. Averaging $\P\B\A(\mpsi)$ over
one period delivered  $\Gone = -3.97,
\Gtwo = 0.70$ for Barkley kinetics and $\Gone = 0.455$,
$\Gtwo=0.302$ for FK kinetics. 
Theoretical predictions \eqref{eom2}, \eqref{eombik} using the measured
  $\G_{1,2}$ are in good agreement with the observed drift 
  of spirals in constant field $\E$, and with circular scroll ring dynamics,
  see Fig.~\ref{fig:tension}.

Since the chosen
parameters in Barkley kinetics yield $\Gone<0$, the filament will
undergo Euler buckling beyond a critical thickness, as we have already seen
Fig.~\ref{fig:scroll}a. The FK model  has $\Gone>0$,
and Fig. \ref{fig:scroll}b shows that a transmural filament indeed
relaxes to the minimal length. 
\begin{figure}[tb]
\raisebox{2.3cm}{a)}\includegraphics[width=0.2 \textwidth]
{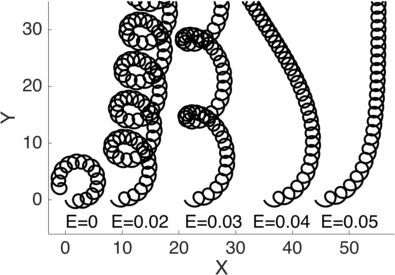}
\raisebox{2.3cm}{b)}\includegraphics[width=0.2 \textwidth]%{vaE.png} \\
{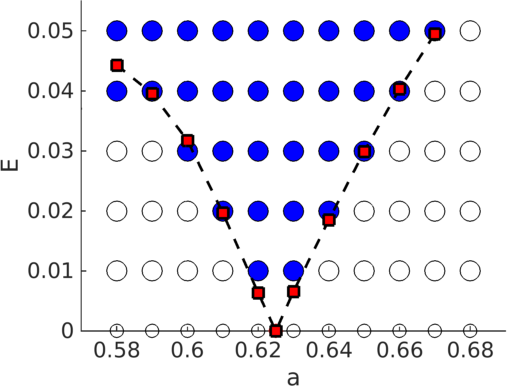} 
\raisebox{2.5cm}{c)}\includegraphics[width=0.22\textwidth]{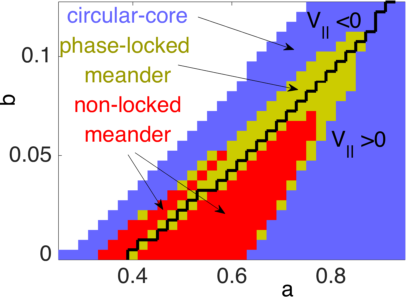} 
\raisebox{2.5cm}{d)}\includegraphics[width=0.22\textwidth]{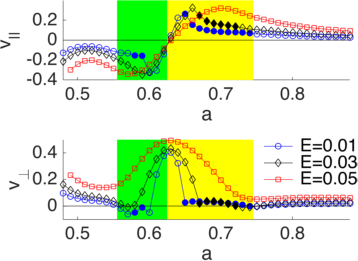} 
\caption[]{\label{fig:PLdrift} %
(Color online) Phase-locking in Barkley's model. (a)  Drift trajectories with
$\vE = \E \ex$ for parameters as in Fig. 1a, showing
phase-locking when $\E>0.04$. (b) Arnold tongue confirming theoretical
prediction in Eq. \eqref{ecrit} (c) Occurence of phase-locking for $\E=0.03$ in (a,b) parameter space with $\cpar = 0.02$. (d) Drift components parallel and perpendicular to $\E$, for $\bpar= 0.05$. Colored background indicates meander.  }
\end{figure}

Until now, it was assumed that perturbations are small and $\romgo$ is not.
As noted already in~\cite{Grill:1996,LeBlanc:2002}, if either condition is
broken, phase-locking between spiral rotation and its meandering may happen.
We are now in a position to describe this phenomenon quantitatively. 
For Barkley
kinetics as in
Fig. \ref{fig:spiral}a where $\romgo=0.08\ll\mOmgo=1.25$, one
finds a qualitatively different tip trajectory when
$\E=|\vE|\geq 0.04$, see Fig. \ref{fig:PLdrift}a. From the first of
Eqs. \eqref{efd_comp2}, one can show 
similarly to \cite{Li:2014} that a necessary condition for locking the rotation phase is 
\begin{align}
  \E > \Ecrit &= \romgo / \Q{}{}, & \Q{}{} =\sqrt{(\avg{\Q\rphi1})^2 + (\avg{\Q\rphi2})^2}.%= 0.041 
    \label{ecrit} 
\end{align}
The locked rotation angle will be
%\begin{align} 
$ \rphil = \arccos\left(- \romgo / \Q{}{} \right) - \arctan\left( \avg{\Q\rphi2} / \avg{\Q\rphi1}\right) $. %\end{align}
Given the computed $\Q\phi\A$,  expression \eqref{ecrit} predicts $\Ecrit = 0.041$, closely matching the value of $0.04$ found in Fig. \ref{fig:PLdrift}a. Fig. \ref{fig:PLdrift}b shows a comparison for different values of parameter $\apar$; it can be seen that the Arnold tongue for phase-locking is well described by Eqs. \eqref{ecrit}. 

In the $(\apar,\bpar)$ parameter space of Barkley's model,
phase-locking is found near the line of resonant meander. Already
  for a field strength of $\E=0.03$, Fig. \ref{fig:PLdrift}c shows phase-locking  in a significant portion of the meander region, where it leads to relatively large drift velocities (see Fig. \ref{fig:PLdrift}d).  In qualitative terms, Fig. \ref{fig:PLdrift}a shows that the meander flower opens up during phase-locking, and the resulting
drift speed is therefore close to the mean `orbital velocity' $\romgo\Rad$ of the tip along the meander flower, where $\Rad$ is the
time-averaged radius of the meander flower. This result does not contradict Eq.  \eqref{efd_comp2} since 
when $\romgo \rightarrow 0$, the centre of the rotating frame is far away. One can instead use a different rotating frame,
  with the origin shifted to the average tip position.  This gives, in leading order,
\begin{align}
  \Vpar(\E) &= \romgo \Rad \cos[\rphil(\E)], &
\Vperp(\E) &= %\V_\y = 
\romgo \Rad \sin[\rphil(\E)]. %+ \OO(\E).
 \label{vlocked}
\end{align}
We noted that the curve
$\Vpar=0$ closely matches the locus of resonant meander. We have
however not found analytical proof of this property, and a
counter-example in the Luo-Rudy-I cardiac tissue model is known
\cite{Alonso:2007}.

\textit{Discussion.} One motivation for this work was to see how the concept of filament tension generalizes to meandering scroll waves. Only when averaged over many meander periods, the dynamics reduces to the circular-core case. The tension concept  has already been used for meander in cardiological literature
\cite{Yamazaki:2012} and modelling studies \cite{Alonso:2007}. Here, we have shown that the emerging property of filament
tension does indeed explain the (in)stability of scroll waves in simple cases, 
  such as those in Fig. \ref{fig:scroll}. Real heart tissue is more complicated in many respects. For instance, a significant
  phenomenon is pinning to heterogeneities. Some pinning effects have  been described before using perturbation methods \cite{Biktashev:2010, Pazo:2004, Biktasheva:2015}. Thin domains of irregular thickness $\Thick(\x,\y)$ (e.g. the cardiac wall) can also be captured by \eqref{efd_comp2}, with $\h = \HP \nabla \ln \Thick$ \cite{Biktasheva:2015}. 

On short time-scales, dynamics is much more complex and the concept of filament tension cannot be applied. The orientation of the meander pattern may phase-lock to external fields, thickness or parameter gradients. 

In general, the theory that was presented here opens
the pathway to analysing and predicting the trajectory and stability
of meandering spiral and scroll waves in reaction-diffusion media of
diverse nature.

HD was funded by FWO-Flanders during part of this work. The computational resources (Stevin Supercomputer Infrastructure) and services used in this work were provided by the VSC (Flemish Supercomputer Center), funded by Ghent University, FWO and the Flemish Government – department EWI. IVB and VNB gratefully acknowledge EPSRC (UK) support via grant 
EP/D074789/1. IVB acknowledges EPSRC (UK) support via grant 
EP/P008690/1. VNB acknowledges EPSRC (UK) current support via grant 
EP/N014391/1 (UK). 

% \References
% \renewcommand{\bibname}{References}
\bibliographystyle{unsrt} 
%\bibliography{references_Hans} 

\cleardoublepage
\appendix

\section{Numerical integration of the reaction-diffusion equations\label{supp:rde}}

We integrate Eq. \eqref{RDE} forward in time by explicit Euler
stepping with time step $\dt$ on a finite differences grid with spatial
resolution $\dx$ and size $\Nx \times \Ny \times \Nz$. Diffusion is
 implemented using finite differences with a 5-point
Laplacian in 2D and 7 points in 3D.

For Barkley kinetics, we used parameter values $\apar=0.58$,
$\bpar=0.05$, $\cpar=0.02$ unless stated otherwise. The model 
has dimensionless space and time units; we use
$\Nx=500$, $\dx=0.1$ and $\dt=0.002375$ in 2D and $\dt=0.0016$ in
3D. The spiral tip was found every 0.1 time units as the intersection
of the isolines $\uvar=0.5$, $\vvar=0.5\apar-\bpar$
\cite{Barkley:1991} using the algorithm in \cite{Fenton:1998}.

With the Fenton-Karma cardiac tissue model \cite{Fenton:1998}, we
selected the guinea pig (GP) set of model parameters
\cite{Fenton:1998}, since it yields a quasi-periodic meandering spiral
with linear core. Notably, the reaction kinetic functions $\bF(\uu)$
are not continuously differentiable with respect to $\uvar$, so the
Jacobian matrix $\bF'(\uuo)$ in Eq. \eqref{defLL} contains singular
contributions. However, in this study we do not directly solve the
linearized equations, and $\bF(\uu)$ can be regularised by approximating Heaviside function in
  its definition with a smooth sigmoidal function, with any required
  accuracy. We do not carry out this limit procedure here, but the predictions of our theory agree well with the outcome of numerical simulations from the
unmodified Fenton-Karma GP model. We used
$\HP = \diag(0.1, 0,0)\,\mm^2/\ms$ as in \cite{Fenton:1998} and we
decreased the spatial grid to $\dx = 0.15\,\mm$ to find
quasi-stationary rotation. A time step of $\dt=0.053\,\ms$ was chosen
in a grid of size $400 \times 400$. The tip line was tracked as the
intersection of the isosurfaces $\uvar=0.5$ and $\dd_\t\uvar=0$, resulting
in the tip trajectory of Fig. \ref{fig:spiral}b with outer radius
9.1mm.

\section{Numerical evaluation of the overlap integrals $\Q\M\A$ \label{supp:maa}}

\begin{figure}[tb] \ \\ \flushleft
a) \\\includegraphics[width=0.47 \textwidth]{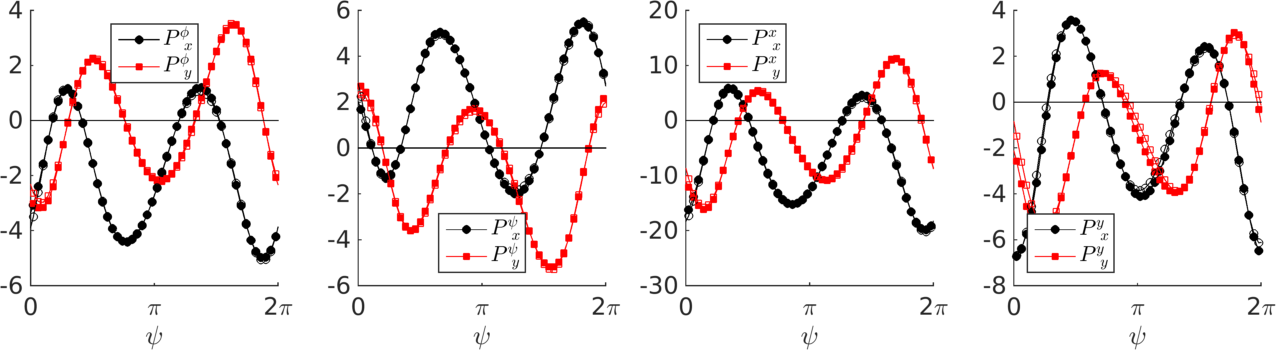} \\
%\raisebox{1.5cm}{b)} 
b) \\
\includegraphics[width=0.47 \textwidth]{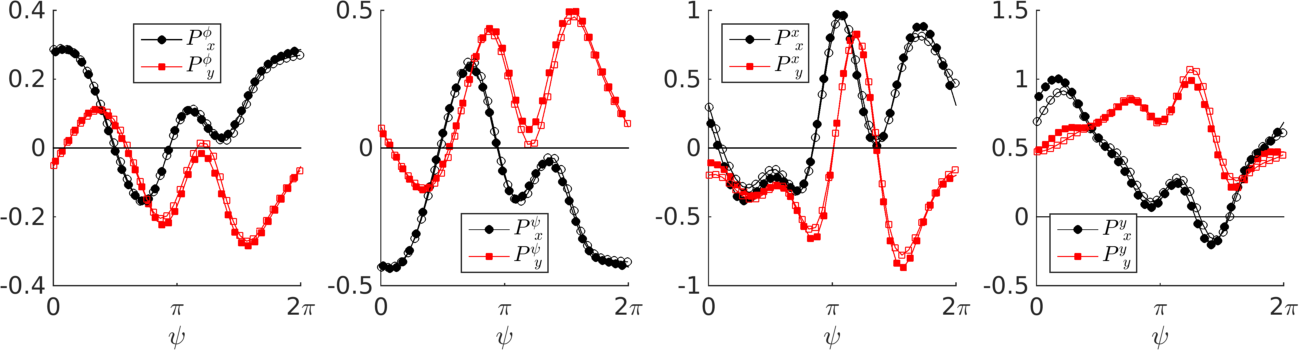} \\
%\raisebox{1.5cm}{a)} 
%a) \\\includegraphics[width=0.47 \textwidth]{fig5a.png}\\ %{matels_2per_Ba.png} \\
%\raisebox{1.5cm}{b)} 
%b) \\
%\includegraphics[width=0.47 \textwidth]{fig5b.png}\\ %{matels_2per_FK.png} \\
\caption{\label{fig:matel} %
  Numerical computation of the matrix elements $\P\m\A$ which
  determine the dynamics of meandering spiral waves in an external
  field with gradient coupling at different meander phases $\mpsi$,
  for (a) Barkley and (b) Fenton-Karma kinetics. The phase $\mpsi=0$
  corresponds with the fiducial points (red dots) in
  Fig. \ref{fig:spiral}. Both panels illustrate the periodicity of matrix elements in the slowly moving frame, by showing them in two subsequent meander cycles (filled and unfilled markers). }
\end{figure}

If the response functions $\WW^\M$ are known with sufficient
accuracy, the matrix elements $\Q\M\N$ can be found by evaluating
the integrals, in a manner similar to the circular-core case
\cite{Henry:2002, Biktasheva:2009}. In practice, however, it is
simpler to measure the  drift induced by an external
field which is imposed during a given fraction of the meander
cycle. First, we measure the absolute position and phase of an
unperturbed meandering spiral wave as detailed in \cite{Dierckx:2017} and denote it as
$\Xo\M = (\Xco, \Yco, \rphio, \mpsio)$. Thereafter, we run many
simulations that deliver a global stimulus at different phases of the
meander cycle: for different $\tptb$
we apply a time-dependent field 
$\vec{\E}(\t) = \Eo\Heav( |\t - \tptb| - \Deltat /2)\,\ex$ 
of duration $\Deltat$, where $\Heav$ is the Heaviside step function. For each simulation, we measure the
absolute spiral phases and position as explained in
\cite{Dierckx:2017} in a time interval $[\t_{start}, \t_{end}]$, where $\t_{start} \gg \tptb$. The length of the interval was taken to be $120 \approx 24\,\To$ for Barkley kinetics and $\ttwo = 1\,\second \approx 40\,\To$ for the FK guinea pig model.  By comparing with a reference case without perturbation, we can compute the matrix elements:
\begin{align}
  \Q\m\xl(\tptb) \approx \frac{\X^\m(\tend) - \X_{\rm ref}^\m(\tend)}{\Eo\Deltat }. 
\end{align}
We repeat the procedure for a field along the $\yl$-direction to find
$\Q\m\y(\tptb)$ and then convert the matrix elements to the
periodic functions $\Q\m\A = \Q\m\a \R\a\A(\rphi)$ which only
depend on $\mpsi$. Note that $2\pi$-periodicity of these functions
is not guaranteed by construction, but is indeed observed with good
accuracy in Fig. \ref{fig:matel}, which we take as an indication that the results are reliable and the analysed spiral wave regimes are indeed biperiodic. 
Fig. \ref{fig:matel} shows the computed curves $\Q\m\A$ for the
Barkley and FK models with parameters as above. Since we took
$\HM = \HP$, the computed coefficients are denoted $\P\m\A$. For
the cardiac tissue FK-model, these results should be interpreted in
terms of filament tension rather than electrophoretic drift, which is
reserved for chemical systems. To find a suitable $\Eo$ we tried the
method first at $\mpsi=0$ for various field strengths and chose the
maximal $\E$ in the model for which the response was still linear in
the field strength. This resulted in 
$\E=0.5$ for Barkley kinetics and and
$\E=0.1\,\mm^{-1}$ for the FK model.

\end{document}